# Coherent addressing of individual neutral atoms in a 3D optical lattice

Yang Wang, Xianli Zhang, Theodore A. Corcovilos[†], Aishwarya Kumar, David S. Weiss*

*Physics Department, The Pennsylvania State University, 104 Davey Lab, University Park, PA, 16802, USA*

*Correspondence to: dsweiss@phys.psu.edu.

†Current address: Department of Physics, Duquesne University, 600 Forbes Ave., 317 Fisher Hall, Pittsburgh, PA 15282, USA.

**Abstract**: We demonstrate arbitrary coherent addressing of individual neutral atoms in a 5×5×5 array formed by an optical lattice. Addressing is accomplished using rapidly reconfigurable crossed laser beams to selectively ac Stark shift target atoms, so that only target atoms are resonant with state-changing microwaves. The effect of these targeted single qubit gates on the quantum information stored in non-targeted atoms is smaller than $3\times10^{-3}$ in state fidelity. This is an important step along the path of converting the scalability promise of neutral atoms into reality.

Trapped neutral atoms possess the essential features of a qubit:[1] their internal states can be initialized and measured, they have long coherence times,[2] and they can be entangled.[3, 4] Since a useful quantum computer needs many qubits, the fact that thousands of neutral atoms can be optically trapped near each other gives them a strong head start over other qubit candidates in terms of scalability. Eigenstate flipping of individual atoms in one and two dimensional arrays have previously been demonstrated, but in situations with no nearby stored quantum information.[5, 6] Coherent addressing of all the atoms in a sparsely populated plane of lattice sites has been demonstrated, in a way that does not affect atoms in the two adjacent planes.[7] A quantum computer requires the ability to arbitrarily change the quantum state of only targeted individual atoms. A high fidelity single qubit gate on a target site in a 2D array of microtraps was recently demonstrated, but without nearby quantum information and with up to 9% state-flipping crosstalk with adjacent atoms.[8] In this paper, we demonstrate coherent addressing of targeted atoms in a 5×5×5 3D array with a minimal effect on nearby quantum information.

Scalable optical traps for individual neutral atoms include 2D optical lattices,[9, 10] arrays of dipole traps, [11-13] and 3D optical lattices.[14] The optimal spacing of individual traps is a balance on the one hand between ready entanglement and high density, which favor closely spaced atoms, and on the other hand independent addressability, which favors more widely spaced atoms. A similar tradeoff exists for 3D vs. 2D geometries, where scaling for entanglement, density and error correction favor 3D, but addressability favors 2D. Whatever the geometry, nearby quantum information must remain unperturbed by the addressing light. We tackle this issue head on, with microwave addressing of individual atoms in a 3D lattice and ac Stark-shifting addressing beams that pass directly through non-target atoms. The methods we demonstrate here can be used to pursue higher densities in any geometry.

Our qubit states are the Cs $F$=3 and 4 $m_F$=0 hyperfine ground sublevels, which we call the storage basis. The addressing scheme uses two circularly polarized addressing beams that cross at the target lattice site (see Fig. 1A), ac Stark shifting the target atom by approximately twice as much as any other atom (see Fig. 1B).[15-17] We address with the storage basis tune-out wavelength (880.250 nm)[18] so the storage states experience minimal force from the addressing beams. The $m_F$=1 (or $m_F$=-1) sublevels constitute the computational basis.[19] A first microwave pulse resonant with the target atom at $\omega_1$ coherently transfers the target atom to the computational basis while leaving all non-target atoms in the storage basis. Then a second pulse at $\omega_2$ can arbitrarily change the target atom's quantum superposition in the computational basis. A third $\omega_1$ pulse then returns the target atom to the storage basis. As described below, a four gate sequence for every two addressed qubits cancels the effects of addressing on non-target atoms.

Our experimental apparatus was largely described in Refs. 14 and 20. A 3D optical lattice is made from three interfering pairs of blue-detuned (847.78 nm) laser beams, with 10$^o$ between the incident angles within a pair, and ~100 MHz frequency offsets between pairs, which creates a 200 μK deep, 4.9 μm spaced lattice in each direction with negligible tunneling. A random ~40% of the lattice sites in our target 5×5×5 site array start with one atom, while the rest are empty.[14] We detect the occupancy of all lattice sites in a plane with negligible error by collecting the fluorescence from polarization gradient cooling with a 0.55 numerical aperture composite lens. By successively translating the lens along the optical axis we make a 5 plane occupancy map (in 850 ms) without affecting the occupancy. The final atom hyperfine state is measured by clearing the $F$=4 atoms with a resonant pulse, then making a new occupancy map. We cool the atoms in the $|F=4, m_F=4\rangle$ magnetic sublevel to near their vibrational ground state with projection sideband cooling.[20] We then use a series of five or six adiabatic rapid passage microwave pulses [20] to transfer them to $|3,0\rangle$ or $|4,0\rangle$ state with 97.2% efficiency, where the loss is due to atoms that are left in high vibrational states after cooling, an issue that can likely be avoided with the use of higher lattice power.[21] After transfer, we adiabatically halve the lattice trapping frequency to 7.5 kHz.

We have a plan for converting partial site filling into unity occupancy in a sub-volume that requires some of the methods demonstrated here.[15, 16] For now, we simply address on a site-selective basis, regardless of the presence of an atom there. The images and data we report are produced from multiple such implementations.

To address a target atom, we use two orthogonal, 2.7 μm waist circularly polarized laser beams with 26 μm Rayleigh ranges (see Fig. 1A). The beams are each reflected from a pair of microelectromechanical systems (MEMS) mirrors, whose angles can be controlled over 0.5 degrees with a ~5 μs reset time.[22] Using a 5-element optical transfer system, the angles are converted into beam translations at the atoms with a 30 μm range.[22] A series of control systems maintains the stability of the addressing beams, which we align to the atoms using atomic signals.[23]

Site selective state transfer is illustrated by Fig. 2. We point the addressing beams to the first target site and then turn on the light in 290 μs. We apply a microwave π-pulse, whose frequency we scan across the range that includes the $|4,0\rangle$ to $|3,-1\rangle$ resonance for all atoms. We then adiabatically turn off the addressing light, move the MEMS mirrors, and repeat the process, sequentially targeting two lattice sites in each of two planes. Fig. 2A plots $R$, the ratio of the number of detected $F$=3 atoms to the initial number of atoms. The green points are the signal from atoms that are never in the line of an addressing beam. The blue points are from atoms that are shifted by only one addressing beam. The orange points correspond to the signals from targeted sites. Fig. 2B shows the summed atom pictures of the two planes with addressed atoms and the plane in between, taken at the target frequency at the orange peak in Fig. 2A. The background 1.7%±0.4% of non-target atoms in the F=3 state are unrelated to the microwave transfer, but rather due to lattice spontaneous emission, and imperfections in the clearing process and the transfer from the |4,4⟩ state.

Fig. 3 illustrates coherence in these lattices with data on the clock transition without addressing beams. We start with the atoms in the |3,0⟩ state, apply a microwave π/2 pulse, wait for a time *T/2*, apply a π-pulse, wait for another T/2, and then apply a final π/2 pulse with a scanned phase. We plot the contrast of the resulting fringe as a function of *T*, with insets showing representative fringes. The echo is needed mostly because of imperfect sideband cooling, which leaves from 25 to 40% of the atoms in higher vibrational levels with different lattice light differential ac Stark shifts (130 Hz/vibrational level); $T_2$ is 26 ms for the central 3×3×3 core, and 10 ms overall. The coherence time ($T_1$) exceeds 7 s, limited by lattice spontaneous emission. The coherence time significantly exceeds those of other single atom neutral atom experiments because these atoms are vibrationally cold in all directions, which minimizes inhomogeneous broadening and spontaneous emission.[3, 24, 25] Coherence times for storing light with trapped atoms of 16 s have been achieved using a "magic" magnetic field;[26] while perhaps applicable here, such large fields would come at the expense of the robustness of our storage basis. There is a ~10% loss of atoms in the detected signal (visible in the Fig. 3 inserts), 3% from imperfect transfer and 7% from a ~10 s collision rate with background gas atoms. Improved vacuum and more lattice power to improve and speed up the projection cooling should ultimately bring losses below 1%.

To demonstrate coherent addressing, we first use a π/2 pulse (at $\omega_0$), to put all the atoms into the superposition (|3,0⟩ + |4,0⟩)/√2. We then execute the addressing procedure discussed above. By empirically adjusting the microwave polarization, we equalize the |3,0⟩ to |4,1⟩ and |4,0⟩ to |3,1⟩ Rabi frequencies, so the transfer to and from the computational basis can be made with a simple π-pulse. We generate the three microwave frequencies with a single direct digital synthesizer. The phase of the $\omega_2$ pulse can be adjusted so that it rotates the Bloch vector about any axis in the **X**-**Y** plane, and its amplitude can be adjusted to give any rotation angle (for example, see Fig. 4B). Thus arbitrary rotations on a Bloch sphere can be made. A final $\omega_0$ π/2 pulse with scanned phase then probes the quantum states of all the atoms.

We have demonstrated single qubit gates on two non-coplanar target atoms, with the MEMS mirrors redirection during the sequence. Each application of an addressing beam results in a ~0.35π phase shift for the non-target atoms on its line, and the microwave pulses cause 0.1π scale off-resonant ac Zeeman phase shifts on non-target atoms. To cancel these unwanted shifts, we use "dummy" gates to ensure that each non-target atom experiences each of these shifts twice, with its state flipped between the two times, as illustrated by the pulse sequence in Fig. 4A. The orange circles in Fig. 4C show results for a π-rotation about the **X**-axis (gate I), in Fig. 4D a π-rotation about the $(\mathbf{X}+\mathbf{Y})/\sqrt{2}$ axis (gate II), and in Fig. 4E a π/2-rotation about the **X**-axis (gate III). The green diamonds in Figs. 4C-E show interference fringes for non-target atoms that see no addressing light, the blue triangles for non-target atoms that are in line with addressing beams, and the pink squares for the 6 atoms adjacent to each target atom.

The spin-echo type approach that we employ on non-target atoms only imperfectly cancels unwanted shifts for the target atoms, since the phase shifts experienced by a target atom during the dummy pulses differ from those during the targeting pulse. Accordingly, we empirically cancel the shifts due to the dummy pulses when adjusting the phase of the $\omega_2$ pulse for each gate. Eventually it will be necessary to model all dummy shifts to predetermine their effect on the gates.

The fringes of Figs. 4 C-E provide the basic information about gate fidelity, allowing us to project the Bloch vector of an atom before the final π/2 pulse onto its target Bloch vector.[23] The spin echo infrastructure compromises the (identity gate) fidelity of non-target atoms by 0.01 to 0.02, depending on how carefully pulse parameters have been adjusted. The addressing light and microwaves cause a marginal change in identity gate fidelity of -.001±.003 and +.002±.003 for the line atoms and nearest neighbors respectively; that is, even the most vulnerable non-target atoms are unaffected at the level of 0.003. The fidelity of the gates on the target atoms themselves (including the spin echo errors) are 0.95, 0.91 and 0.93 for the gates in Figs. 4 C, D and E respectively. Gate I has higher quality because the two coherent halves of the atom spend the same time in all sublevels. Gate fidelities can ultimately be improved by varying the spin echo pulse phases,[27] better projection cooling, better B-field stability, more tightly locked addressing beam positions, better addressing beam spatial modes, and quantum control techniques.[28]

Given the 7 s coherence time and the ~1 ms total time for the each single qubit gate, we can in principle perform thousands of such gates before significant decoherence occurs. Deeper, farther detuned lattices will ultimately allow for superior and faster initial cooling, as well as faster gates. The next steps on the road to a quantum computer in this system will be to fill in vacancies and implement entangling Rydberg gates in this already highly scaled system.

We thank the DARPA QuEST program and the Army Research Office for supporting this work.

19 It is also possible to use the $|3,1\rangle$ and $|4,-1\rangle$ states for storage, and our storage states for computation, since both manifolds are first order B-field insensitive, but that choice

# Figures

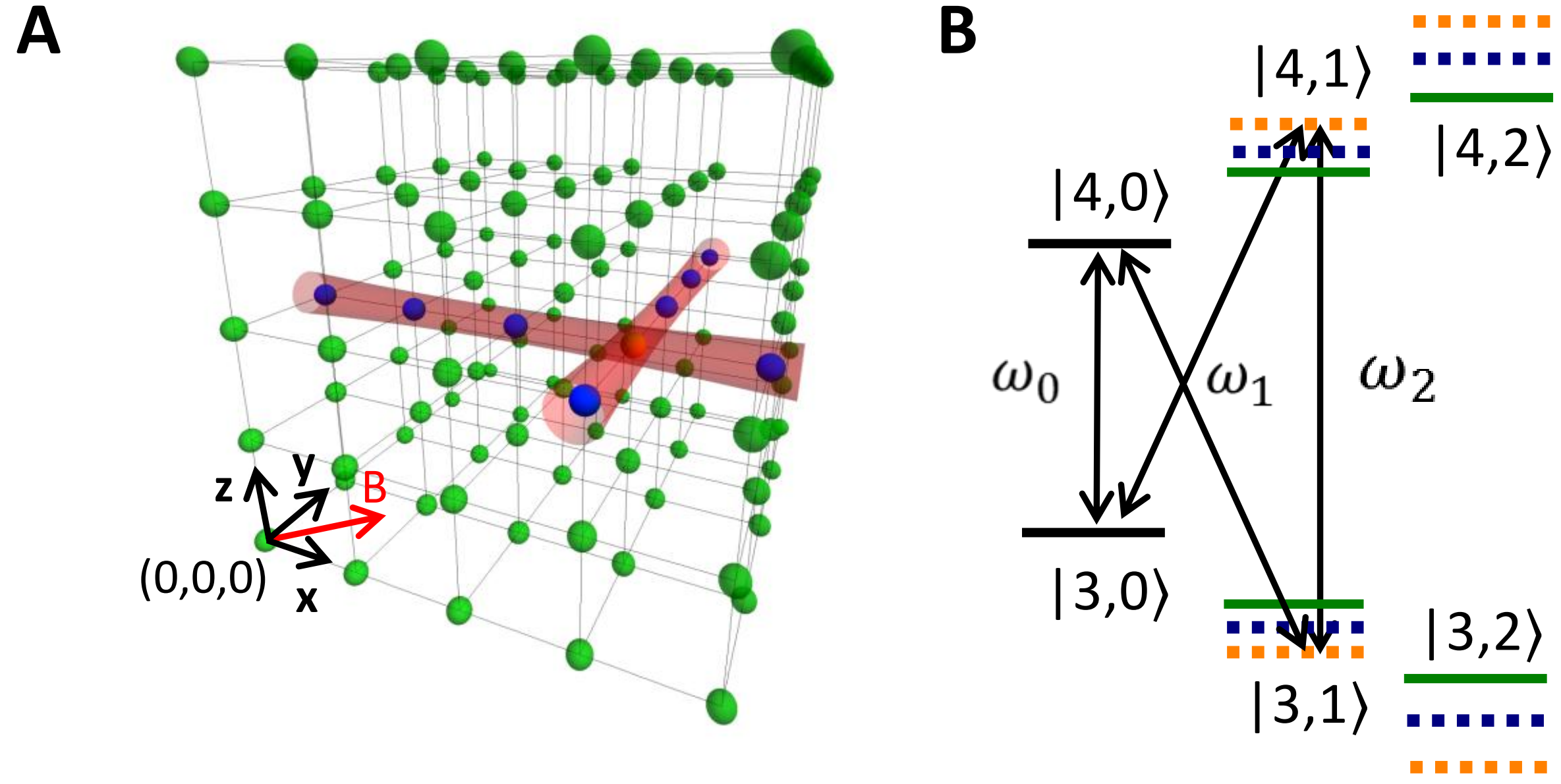

**Fig. 1: A.) Diagram of addressing a 5×5×5 array of neutral atoms. Each addressing beam can be parallel-translated within 5 μs to any line of atoms, so that any site can be put at their intersection. The addressing beams are circularly polarized, and the 140 mG magnetic field is in the same plane. B.) The relevant part of the ground state energy level structure for addressing (not to scale.) A target atom experiences twice the ac Stark shift of any other atom (its shift is illustrated by the orange dashed lines), so that, starting in the storage basis, |3,0⟩ and |4,0⟩, it alone is resonant with ω1. After it is transferred to the computation basis, |3,1⟩ and |4,1⟩, it alone is resonant with ω2.**

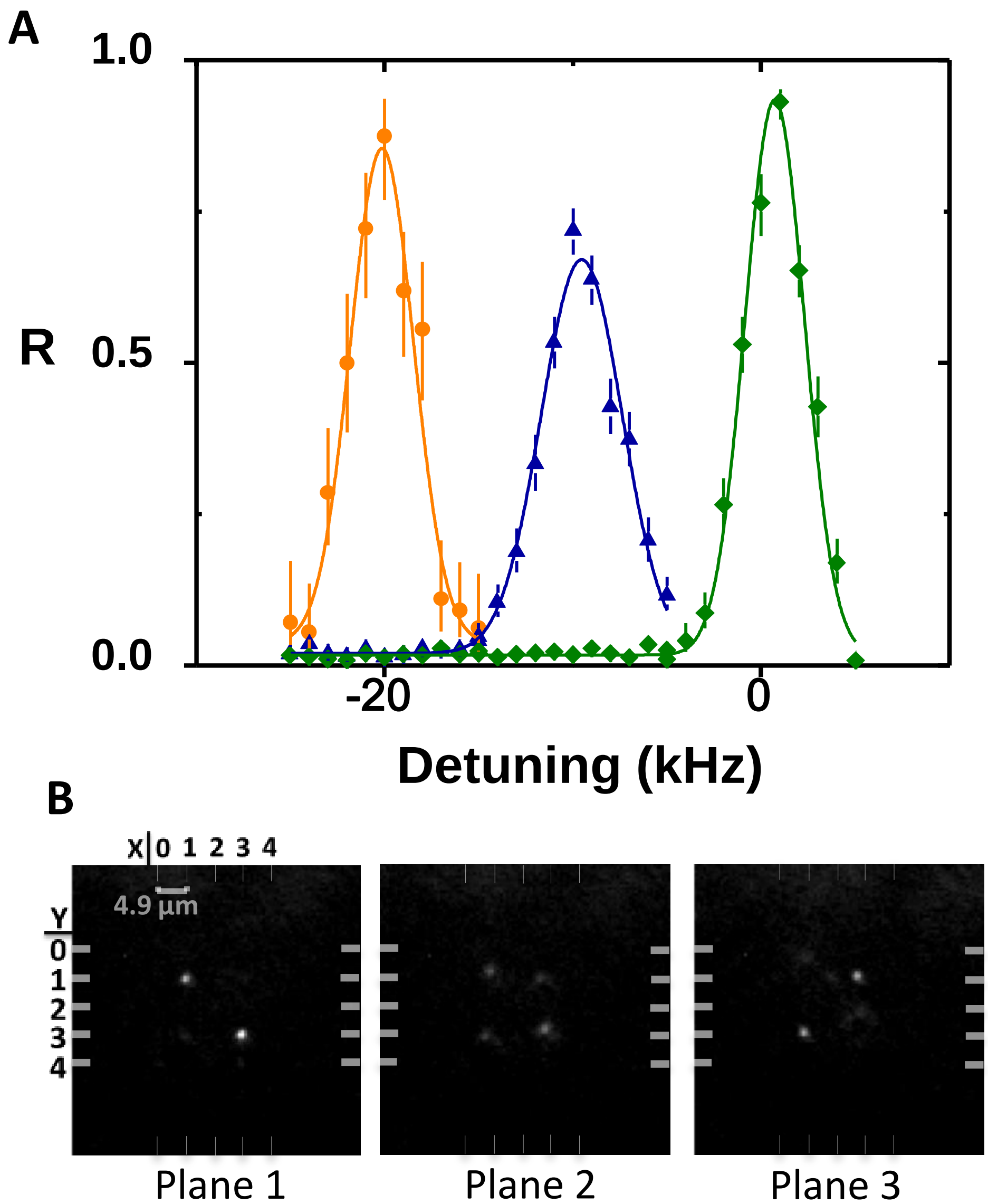

**Fig. 2: A.) Probability that an atom makes a transition from |4,0⟩ to |3,-1⟩ as a function of the detuning of ω1 from the unshifted resonance. The orange points correspond to the four target sites in two planes addressed in this sequence, the blue points correspond to non-target atoms traversed by addressing beams, and the green points to all the other atoms. The solid lines are fits to Gaussians. The error bars are from counting noise. B.) Summed images of the raw signals from three planes using the value of ω1 that yields the orange peak in A. Two sites in each of Planes 1 and 3 are addressed in each implementation (labeled by (1,1,1),(3,3,1), (1,3,3) and (3,1,3)). The signal in Plane 2 is the out-of-focus light from the adjacent planes; out-of-focus atom images have nearly double the radius, the probability of them giving a false positive in our occupancy maps is below $10^{-6}$.**

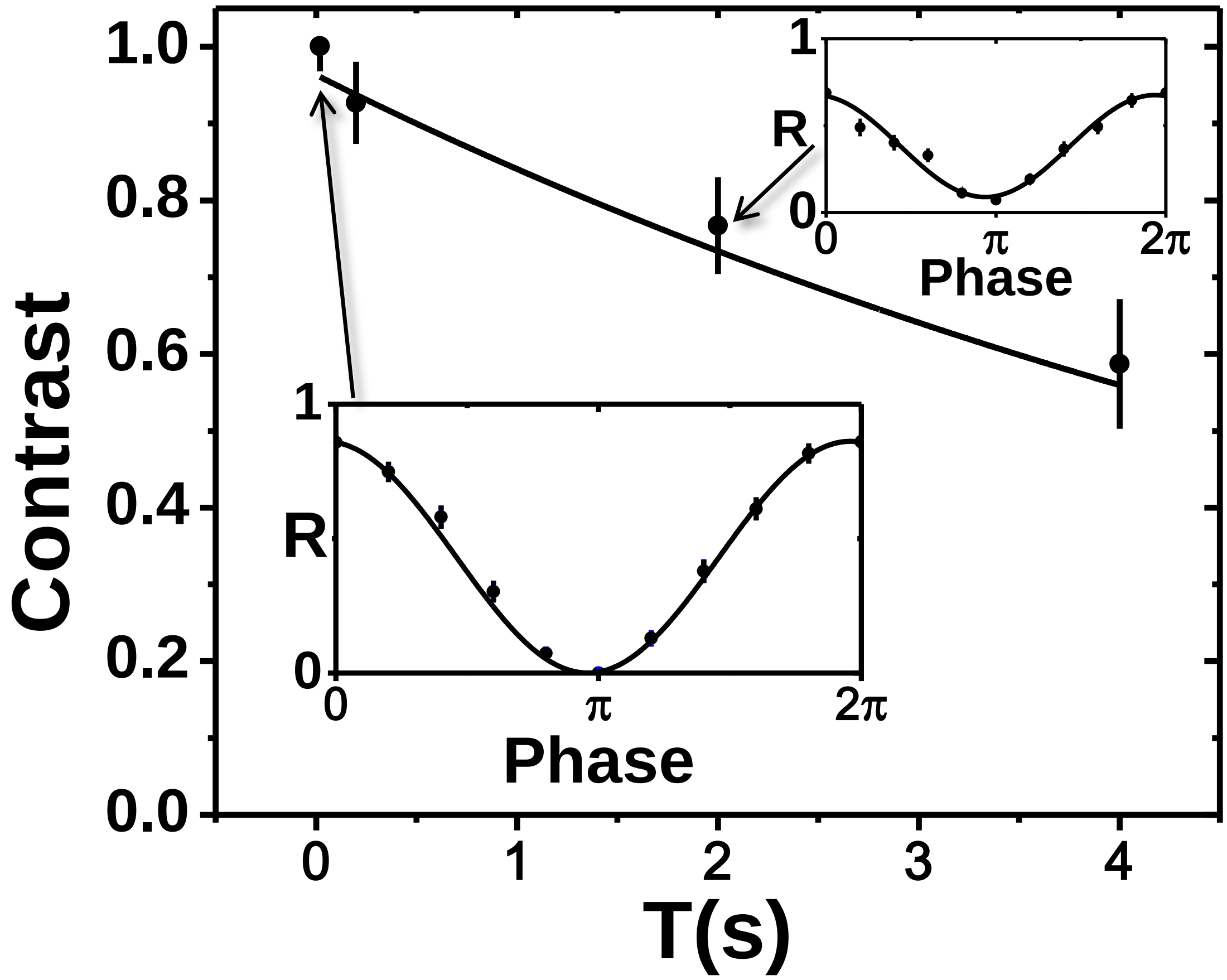

**Fig. 3: Fringe contrast for a spin-echo sequence on all the atoms. The fit exponential time constant is 7.4 s. Since the contrast is lost due to spontaneous emission, the rate of which depends on an atom's vibrational state, the true function is more complicated. The insets are the unnormalized fringes at the indicated times, where the phase of the final π/2 pulse in a π/2-π-π/2 is varied.**

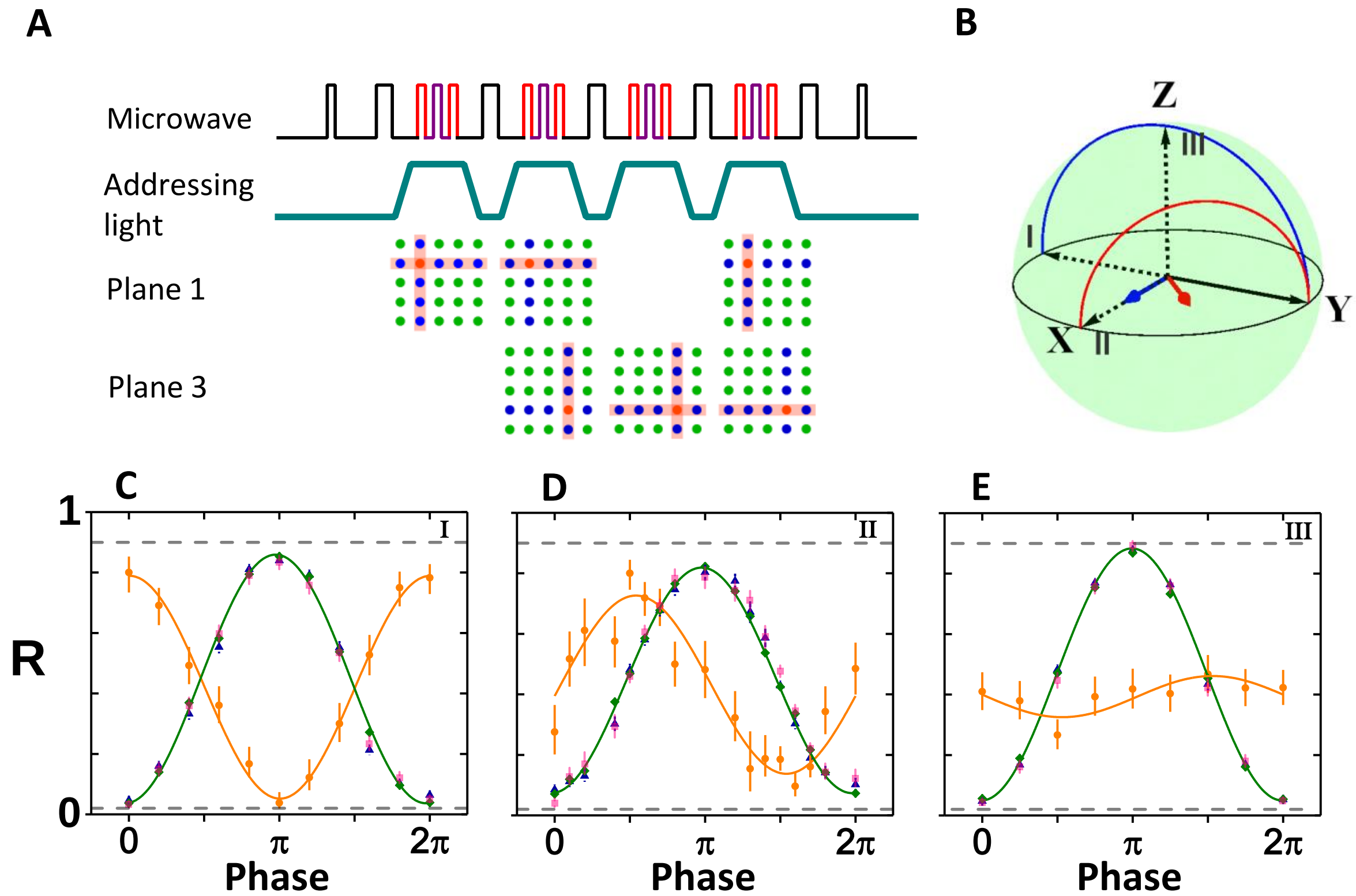

**Fig. 4: Arbitrary single qubit gates. A.) The addressing pulse sequence. The top row shows the timing of the Blackman profiles microwave pulses. Black corresponds to ω0, red to ω1, and purple to ω2. The timing gaps are enhanced for clarity. The second row shows the addressing beam pulses vs. time, and the third and fourth rows show the corresponding addressing beam locations. The atom color scheme is that of Figure 1. B.) A Bloch sphere representation of qubit rotations due to the ω2 pulses. The solid black arrow is the initial quantum state, the blue and red arrows correspond to torque vectors, the red and blue arcs are the Bloch vector paths during the gates, and the dashed black arrows are the final states, labeled I, II and III. C.) Interference fringe due to gate I of Fig. 4B. The orange circles are due to the four target atoms, the blue triangles are for line atoms, the pink squares are for nearest neighbors, and the green diamonds are for other atoms. The gate target is a π-shifted fringe. D.) Interference fringe due gate II of Fig. 4B. The gate target is a π/2-shifted fringe. E.) Interference fringe due to gate III of Fig. 4B. The gate target is a horizontal line of half the peak of the unaffected fringe.**

## Coherent addressing of individual neutral atoms in a 3D optical lattice

Yang Wang, Xianli Zhang, Theodore A. Corcovilos[†], Aishwarya Kumar, David S. Weiss*

*Physics Department, The Pennsylvania State University, 104 Davey Lab, University Park, PA, 16802, USA*

# Supplementary material

## Stabilization of the system

There are three optical systems, the imaging system, the lattice system, and the addressing system, that must all be aligned with respect to each other and the alignment must be maintained throughout an extended data run. To accomplish this, we must correct changes in the pointing and phase of these systems due to slow thermal drifts in the laboratory.

### *Imaging objective lens position feedback*

The reference position for all three systems is the center of the field of view of the imaging system. The imaging objective lens is mounted on a precision translation stage on the optical table. A strain gauge monitors the position of the lens within the stage with an uncertainty below 10 nm, and the reading is fed back to a piezoelectric stack to maintain a fixed position. The position set point is dynamically changed during each experimental run in steps equal to the z lattice spacing (4.9 μm) to image each of the several lattice planes used in the experiment.

### *Lattice phase feedback*

The location of each lattice site is determined by the interference patterns generated by 3 pairs of phase coherent beams, where the beam directions within each pair differ by approximately 10°. The optical path lengths of each beam in a pair are matched in order to minimize the sensitivity of the interference pattern to changes in the lattice laser wavelength. A phase shift in one beam of a pair results in a translation of the interference pattern. We have observed slow drifts in the position of the interference pattern with a characteristic time scale of <1 μm/hour, which we attribute to thermal changes in the optical path lengths of the beams. We correct for this drift using a Brewster's angle plate (0.3 mm thick, undoped YAG crystal, index of refraction = 1.82) in one beam of each lattice pair, which we tilt with a kinematic mount driven by a linear piezo stepper motor. The tilt produces a phase shift in one leg of a pair. A tilt of 8 mrad yields a phase shift of $\pi/2$ and a shift in position of 4.9um. After each iteration of the experiment, we calculate the absolute position of the interference pattern, as described below, and apply a small correction to each Brewster plate angle. The time constant used in the correction is 120 s.

To determine the absolute position of the lattice, we compare the experimental images of the atoms to a previously measured 3D reconstruction of the imaging system point-spread function. From each experimental image set, we identify atoms that have no nearest neighbors along the imaging axis. The images of only these isolated atoms are averaged together, along with the images of the vacancies immediately above and below the atom which show a defocused image of the isolated atoms. These three averaged images, one in front of the target atom, one centered on the target atom, and one behind the target atom, are fit to the 3D point spread function of the imaging system to give the absolute 3D position of the lattice relative to the imaging system. This position measurement becomes the input for a discretely stepped software PID loop, which then makes a one-time adjustment to the position of each Brewster plate. After each experimental iteration, the PID loop and lattice position are updated using the most recently acquired image data.

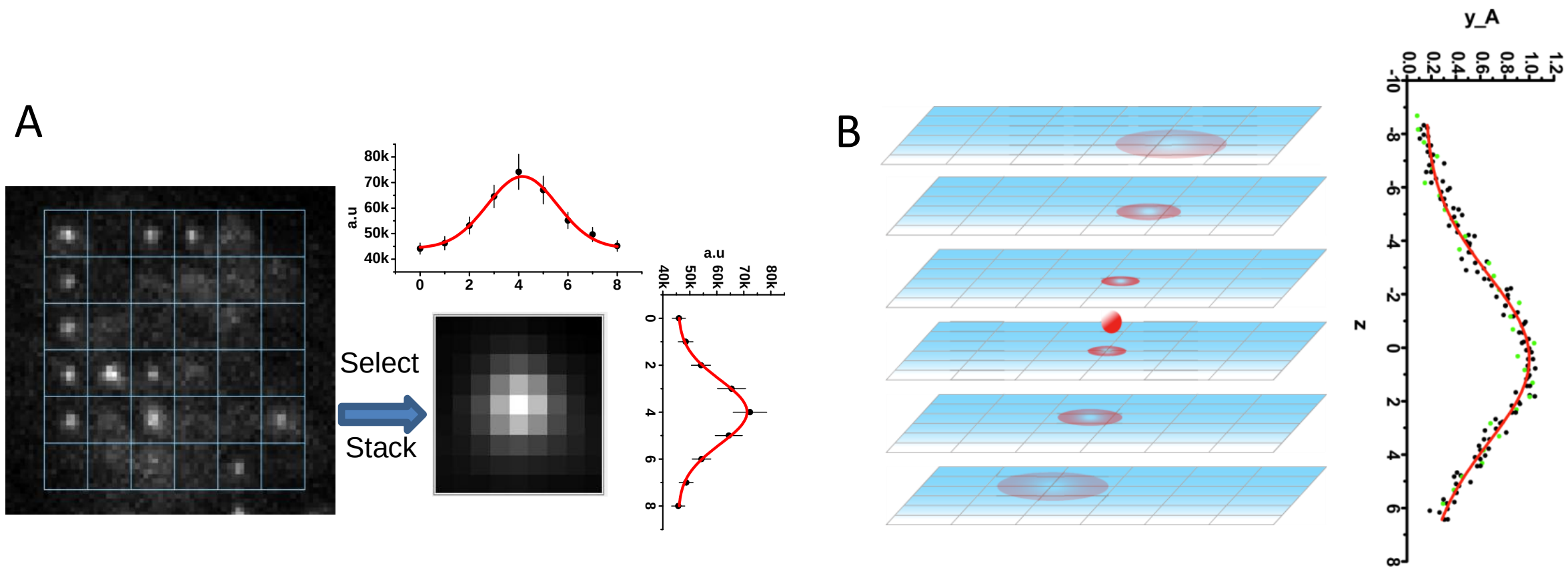

**Fig. 1: Lattice position measurement. A.) Raw and stacked images of atoms in the in-focus plane. Only the atoms with no neighbors along the imaging axis are selected for the stacked image. X and Y amplitudes are plotted and fit to a Gaussian to give X and Y position of the lattice. Similar stacked images are obtained for atoms one lattice site out of focus. B.) Z-calibration. The intensity amplitude from the Gaussian fit of the stacked images is plotted against the Z coordinate. The peak intensities from the three stacked images, i.e., in-focus, near-defocused and far-defocused, are fit to determine the Z error.**

With the lattice feedback system in place, the absolute position of the lattice can be maintained to within 0.1um in the imaging plane and 0.23um along the imaging axis. The stability is limited by the uncertainty of the position measurement, particularly along the imaging axis.

## Alignment of addressing beams

The addressing beams need to be well-aligned with atoms for them to be maximally insensitive to position fluctuations. To align the beam to the atom arrays we use microwaves to drive transitions between hyperfine states (see Fig. 2A in the main text) while stepping through transverse positions. We then use the population of atoms to indicate how far away the beam is from the atoms. The frequency of the microwave is chosen to be just above resonance for an atom that is maximally shifted by one beam. As the beam is scanned spatially, the most atoms are transferred when the alignment is best. We perform this alignment on the two transverse directions of each beam sequentially, and iterate as necessary. With this method we can achieve 100 nm alignment precision.

## Gate Fidelity Calculation

We calculate the fidelity of the gates for each class of atoms (target, line, nearest neighbors and spectators) by reconstructing the state of the atoms before the final π/2 pulse in the spin echo gate sequence and overlapping it with the expected state after a perfect gate operation. The state is reconstructed by fitting the spin echo data shown in Figure.4 C-E to the expected functional form. We first normalize the data to take into account the 10% loss from the sum of background gas collisions and the imperfect transfer to storage states, and the 2% population leakage to the F=3 state, half from the imperfect clearing process and half from spontaneously emitted lattice light. Before the final π/2 measurement pulse the state can be expressed as:

$$|\psi\rangle = n cos\left(\frac{\theta}{2}\right)|0\rangle + ne^{i\phi} sin\left(\frac{\theta}{2}\right)|1\rangle.$$

In this equation, $(\theta, \phi)$ follows the standard definition of a single qubit state on a Bloch sphere, $n$ represents a spherically symmetric shrinkage of the Bloch sphere, and $|0\rangle$ and $|1\rangle$ correspond to F=3, $m_F = 0$ and F=4, $m_F = 0$ respectively. The shrinkage comes from any process related to the gate that transfers population out of storage basis. We measure the population in $|0\rangle$. After the final π/2 pulse, whose phase α is scanned, the population is given by,

$$P_0 = \frac{n^2(1 + \sin(\theta)\cos(\alpha + \phi))}{2}$$

After fitting the data to this function, the values for $(n, \theta, \phi)$ and thus the Bloch vector before the final π/2 pulse can be obtained. The density matrix $\rho = |\psi\rangle\langle\psi|$ can be constructed from the Bloch vector. If σ is the expected density matrix for a particular class of atoms, then the fidelity of the gate is given by,

$$F = Tr\left(\sqrt{\sqrt{\rho}\sigma\sqrt{\rho}}\right)$$

| | π rotation around **X** | π rotation around **(X+Y)**/√2 | π/2 rotation around **X** |
|---|---|---|---|
| Spectator | 0.988±0.002 | 0.978±0.002 | 0.990±0.001 |
| Line | 0.984±0.007 | 0.977±0.005 | 0.992±0.003 |
| Target | 0.946± 0.008 | 0.913± 0.023 | 0.925±0.047 |
| Nearest Neighbors | 0.983±0.004 | 0.985±0.007 | 0.993±0.004 |

**Table 1: Fidelities for all the classes of atoms for each of the operations implemented.**

Table 1 shows the fidelities for various classes of atoms each of the 3 different gate operations we implement. The fidelities of the spectator atoms are statistically indistinguishable from that of the line and nearest neighbor atoms. The variance from gate to gate reflects how soon before the data was taken the microwave pulse amplitudes, which can drift, were adjusted. The vast majority of the gate fidelity loss at non-target sites comes from imperfections in the spin echo infrastructure, not the addressing operations themselves. In the future, we will employ standard NMR techniques to drastically reduce our sensitivity to microwave amplitudes.

It is worth noting that this method does not take into account non-spherically symmetric distortions of Bloch sphere. Doing so would require a full characterization of the operations using process tomography, a process that does not seem warranted while there are still significant gate improvements to be made. Still, we think we have demonstrated the highest fidelity operation ($\pi$ rotation around **X**) and the worst fidelity operations ($\pi$ rotation around $(\mathbf{X}+\mathbf{Y})/\sqrt{2}$) for this kind of gate.